\documentclass[12pt]{iopart}
\usepackage{graphicx}
\usepackage{amssymb}
\usepackage{psfrag}
\usepackage{url}
\usepackage{epstopdf}
\usepackage{color}
\usepackage{bm}
\usepackage{textpos}
\def\fun#1#2{\lower3.6pt\vbox{\baselineskip0pt\lineskip.9pt
\ialign{$\mathsurround=0pt#1\hfil##\hfil$\crcr#2\crcr\sim\crcr}}}

\definecolor{Black}{named}{Black}
\definecolor{Green}{named}{Black}
\definecolor{Red}{named}{Red}

\newcommand{\green}[1]{\color{Green} #1\color{Black}}
\newcommand{\yellow}[1]{\color{Black} #1\color{Black}}

\newcommand{\J}[1]{\color{black} #1 \color{black}}

\newcommand{\fex}{{\it e.g.}}

\newcommand{\GeV}{\,\textrm{GeV}}
\newcommand{\TeV}{\,\textrm{TeV}}
\newcommand{\cm}{\,\textrm{cm}}

\newcommand{\s}{\,\textrm{s}}

\newcommand{\kpc}{\,\textrm{kpc}}

\begin{document}

\title[Constraining Decaying Dark Matter with Fermi LAT
Gamma-rays]{Constraining Decaying Dark Matter with Fermi LAT Gamma-rays}
\author{Le Zhang$^1$, Christoph Weniger$^2$, Luca Maccione$^2$, Javier
Redondo$^3$ and G\"unter Sigl$^1$} 
\address{$^1$ II. Institut f\"ur theoretische Physik, Universit\"at Hamburg,
Luruper Chaussee 149, D-22761 Hamburg, Germany} 
\address{$^2$ Deutsches Elektronen-Synchrotron, Notkestra\ss e 85, D-22607
Hamburg, Germany}
\address{$^3$ Max-Planck-Institut (Werner Heisenberg Institut), F\"ohringer
Ring 6, D-80805 M\"unchen, Germany}

\begin{abstract}
 \J{High energy electrons and positrons from decaying dark matter can produce a significant
 flux of gamma rays by inverse Compton off low energy photons in the interstellar radiation field. 
 This possibility is inevitably related with the dark matter interpretation of the observed PAMELA and FERMI excesses.
 The aim of this paper is providing a simple and universal method to constraint dark matter models which produce electrons and positrons in their  decay by using the Fermi LAT gamma-ray observations in the energy range between 0.5 GeV and 300 GeV.
We provide a set of universal response functions that, once convolved with a specific dark matter model produce the desired constraint. 
Our response functions contain all the astrophysical inputs such as the electron propagation in the galaxy, the dark matter profile, the gamma-ray fluxes of known origin, and the Fermi LAT data.
 We study the uncertainties in the determination of the response functions
 and apply them to place constraints on some specific dark matter decay
 models that can well fit the positron and electron fluxes observed by
 PAMELA and Fermi LAT. To this end we also take into account prompt
 radiation from the dark matter decay. 
 We find that with the available data decaying dark matter
 cannot be excluded as source of the PAMELA positron excess.}
\end{abstract}


\begin{textblock}{3}(9.2,-10) 
\noindent 
DESY 09-220 \\ MPP-2009-222 
\end{textblock}
\maketitle

\section{Introduction}
The existence of dark matter in our Universe is nowadays a widely accepted
fact. Dark matter constitutes the dominant fraction of matter present in the
Universe. However, all the astrophysical evidences for dark matter, such as
those coming from gravitational lensing, galaxy rotation curves and cosmic microwave background (CMB) anisotropies, are purely
gravitational~\cite{Bertone:2004pz}, and its particle nature remains unknown.

The most popular type of dark matter (DM) candidate, a weakly interacting
massive particle (WIMP), can naturally reproduce the observed DM abundance due
to effective self-annihilation in the early Universe. Today, this same
annihilation process could produce an observable contribution to the measured
cosmic-ray fluxes on Earth~\cite{Ullio:2002pj, Pieri:2009je, Regis:2009md,
SiegalGaskins:2008ge} and have impact on cosmological
observations~\cite{Zhang:2006fr}. However, such an indirect detection of DM is
also possible if DM \textit{decays} at a sufficiently large rate. 
Indeed, there exists a number of interesting and theoretically well motivated DM models that
\textit{predict} the decay of DM on cosmological time scales, namely with
lifetimes around and above $\tau_\chi\simeq \mathcal{O}(10^{26}\s)$, which are
typically required to be not in conflict with current observational
limits~\cite{Chen:2003gz, Zhang:2007zzh, Gong:2008gi}. 
Among these candidates is the gravitino in small $R$-parity breaking scenarios (motivated by requiring a consistent thermal history of the
Universe~\cite{Buchmuller:2007ui}) and models of sterile neutrinos, whose long
lifetime is due to tiny Yukawa couplings (see Ref.~\cite{Boyarsky:2009ix} and
references therein). Other interesting models include kinetically mixed hidden
gauge bosons and gauginos~\cite{Chen:2008yi,
Ibarra:2009bm}. Even in models where the DM candidate is stable in the first
place, the consideration of higher-dimensional operators often renders the DM
candidate particle unstable with cosmological
lifetimes~\cite{Arvanitaki:2008hq}. Since the indirect detection signals from
decay differ in general from the ones of annihilation, a dedicated study of
decaying DM signals is \J{warranted}.

On the observational side, recent results by several Cosmic Ray (CR) and
$\gamma$-ray experiments have seriously challenged the standard description
and interpretation of the generation of high-energy CR electrons, positrons
and $\gamma$-rays, thus giving room also to DM interpretations.  In
particular, the PAMELA satellite reported the presence of an increase in the
positron fraction $J(e^{+})/J(e^{+}+e^{-})$ above 10
GeV~\cite{Adriani:2008zr}. If there is no primary source of CR positrons in
the Galaxy, and since secondary positrons originate from the interactions
between CR nuclei (mainly protons and Helium) and the interstellar gas, their
spectrum on Earth should be softer than the primary electron spectrum.
Therefore, the positron fraction is expected to decrease with increasing
energy. The observed rise at high energy (the so called positron fraction
\textit{excess}) leads then to postulate the existence of an additional
primary source of positrons with energies ranging from 10 to at least 100 GeV.
Although considering more carefully standard astrophysical sources, such as
pulsars, provides a possible explanation for this excess~\cite{Hooper:2008kg,
Yuksel:2008rf, Malyshev:2009tw, Profumo:2008ms, Kobayashi:2003kp}, also dark
matter interpretations are possible. In particular, DM decays might provide
the required amount of primary positrons~\cite{Arvanitaki:2008hq, Chen:2008yi,
Ibarra:2009dr, Nardi:2008ix, Ibarra:2008jk, Chen:2009uq}. 
\yellow{However, we should stress that the requirement of an additional primary
positron source is not strictly necessary to interpret the data. The positron
excess can be interpreted also as due to propagation effects, either in the
standard sources or in the ISM. For example, the excess could be due to
inhomogeneities of CR sources~\cite{0902.0376},
or it could be a consequence of reacceleration of secondary cosmic rays in
supernova remnants~\cite{0903.2794}. Also 
an increase of starlight and interstellar gas densities 
could reproduce the positron excess~\cite{0908.1094}.  Lastly,
also nested leaky box models can in principle provide a good fit to the
positron fraction~\cite{0905.2136,0907.1686}.

Apart from the positron fraction, also }the $e^-+e^+$ spectra from the Fermi
LAT measurement~\cite{Abdo:2009zk} and HESS~\cite{Collaboration:2008aaa} are
intriguing. The Fermi LAT instrument measured a total $e^{+} + e^{-}$
spectrum at high energies from 20 GeV to 1 TeV on Earth significantly harder
than expected according to propagation models, and with a hint of the presence
of a broad bump at a few hundred GeV, which is smoothly connected with HESS
observations around the TeV (see, e.g.~\cite{Grasso:2009ma,DiBernardo:2009iu}
for a review of possible interpretations of these observations). This also
provides a motivation to explain the excess using decaying dark matter which
can easily produce large numbers of energetic positrons and electrons.
\J{Both these striking observations have motivated a plethora of works on both
building and constraining new dark matter models}.

High energy CR electrons and positrons produce $\gamma$-rays through the
processes of inverse Compton scattering (ICS) off low energy photons in the
interstellar radiation field (ISRF) and through bremsstrahlung emission due to
the interaction with the interstellar medium (ISM). 
The Fermi Gamma-Ray Space Telescope is currently observing the $\gamma$-ray
sky in the energy range between 30 MeV and 300 GeV with unprecedented angular
resolution and precise energy sensitivity. Data obtained during the first year
of operation of Fermi are about to be released officially, and preliminary
data on the $\gamma$-ray spectrum in some regions of the sky were already
\J{made public} in several Conferences. It is then timely to provide tools to exploit
this detailed information in an efficient way. 

At present, most attempts to construct dark matter candidates as the
interpretation of the PAMELA and $\gamma$-ray data are inevitably model
dependent. Thus, it is useful to develop model independent analysis tools such
as response functions. These are functions of injected electron or positron
energy which are obtained from comparing the $\gamma$-ray spectrum resulting
on Earth after propagation \J{(which we call signal)} with the observed $\gamma$-ray spectrum \J{(which henceforth will be called background)}.  Indeed,
this method has already been used to constrain decaying DM scenarios with the
observed galactic synchrotron radiation maps~\cite{Zhang:2009pr}.  Since the
propagation equation is linear in the source term, each injected electron
energy evolves independently. Therefore, with a finite number of numerical
simulations at different injection energies we can construct a numerical
$e^\pm$-response function of \J{signal-to-background}. The $e^\pm$-response
function encodes our knowledge about the propagation of CR electrons in the
Galaxy, as well as about the distribution of DM, but is independent of the
particle physics model leading to the actual spectra of electrons and
positrons after DM decay. Moreover, the $e^\pm$-response functions we compute
in this work are normalized to the present $\gamma$-ray data. 
Hence, constraints can be obtained by \J{the convolution of} the $e^\pm$-response functions with a
given DM decay spectrum and requiring that the result be smaller than the
product of DM mass and lifetime in suitable units. 
If this condition is violated, the DM model predicts $\gamma$-ray fluxes (from ICS and
bremsstrahlung) that are higher than the data by at least $2\sigma$.

In addition to $\gamma$-rays produced by ICS and bremsstrahlung of pairs
resulting from DM decay, \J{the prompt radiation of $\gamma$-rays during the DM decay can be a very relevant signature of DM decays}. 
However, since prompt radiation depends on the particle physics, we \J{cannot} include it in our
$e^\pm$-response functions and \J{has to be included by hand when pursuing constraints on a specific model.}

The paper is organized as follows. In Sec.~2, we describe our method of
$e^\pm$-response functions to provide constrains on any dark matter
model involving electrons and positrons in its decay products. 
In Sect. 3 we present the $e^\pm$-response functions based on the Fermi data and discuss how
they are improved by the removal of astrophysical contributions \J{of known origin} to the $\gamma-$ray signal. 
In Sect. 4 we apply our $e^\pm$-response functions to some specific dark matter decay scenarios and compare it to the contribution
of prompt radiation to the $\gamma-$ray constraints.  
Finally, in Set. 5 we present our final discussions and summarize.


%
%

\section{Gamma-rays from Decaying Dark Matter}\label{sec:res-gamma}
\begin{figure}[tbp]
\centering
\includegraphics[scale = 0.3]{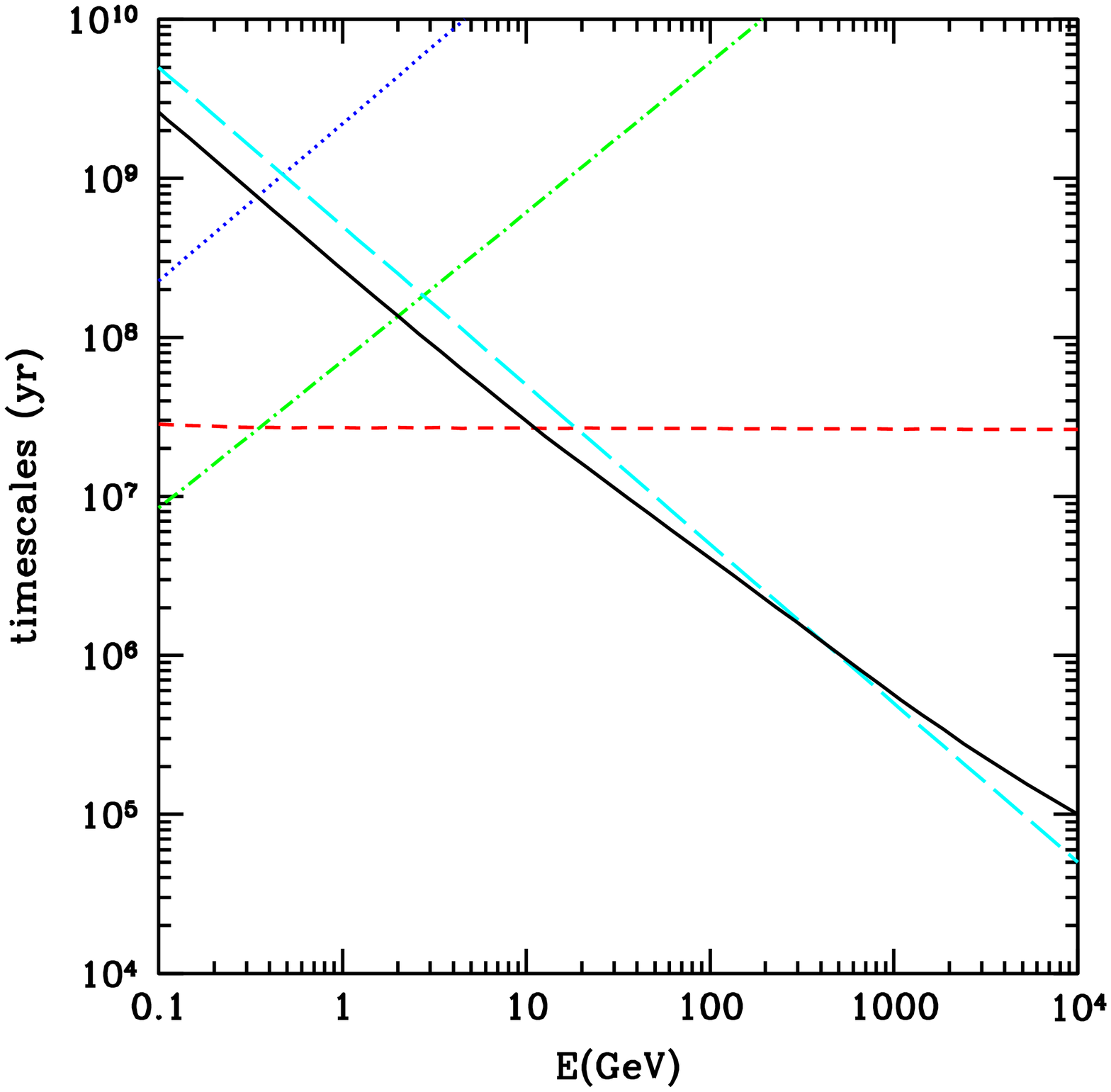}\hspace{0.5cm}
\includegraphics[scale = 0.3]{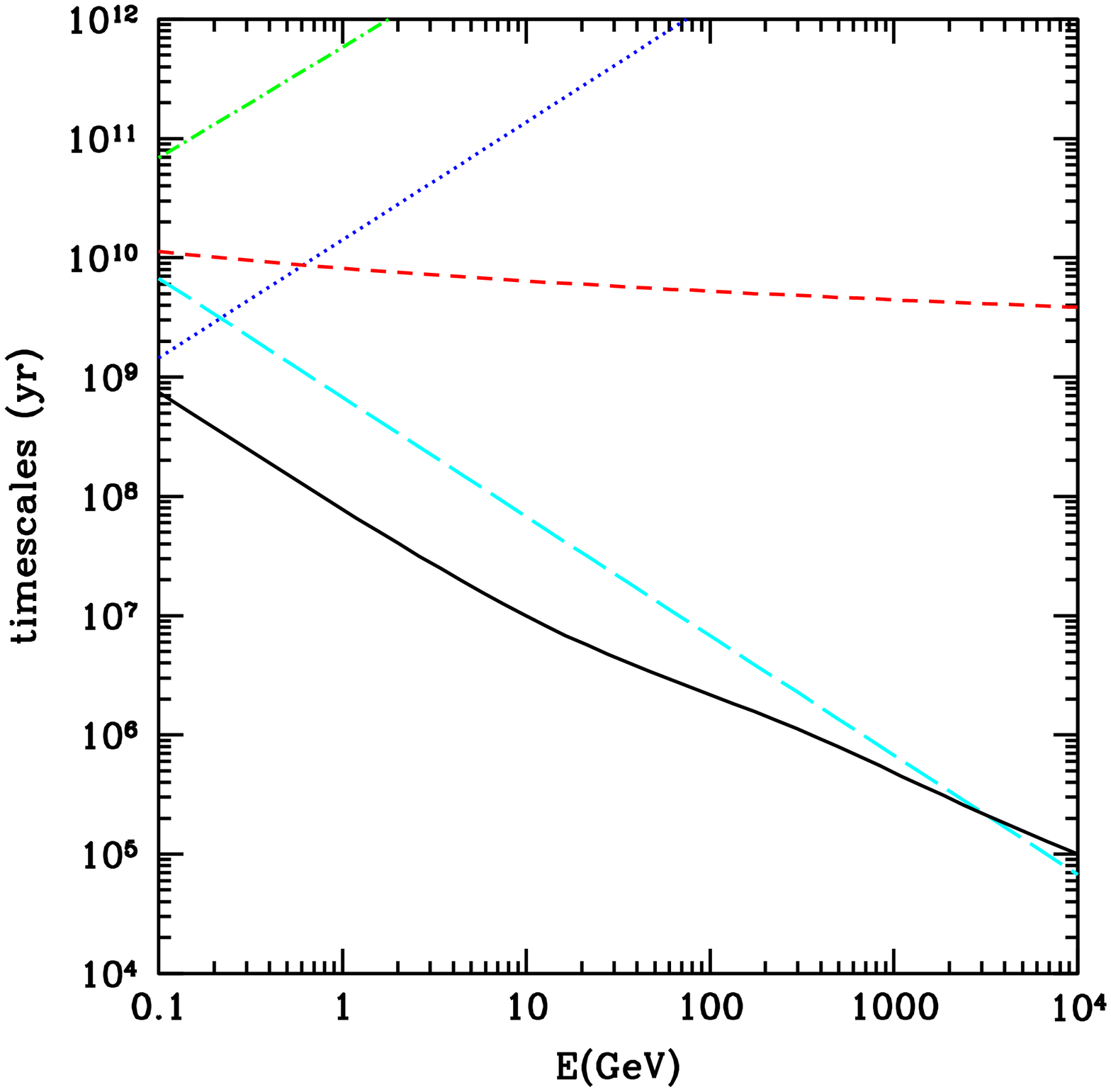}
\caption{Energy loss time scales for CR electrons at Solar position
(\textit{left panel}) and at $r=0$ and $z=2\kpc$ (\textit{right panel}). The
black solid curve represents IC losses, cyan long dashed is for synchrotron
losses, red short dashed for bremsstrahlung losses, blue dotted for Coulomb
losses, green dot-dashed for ionization losses.}
\label{fig:energylosses}
\end{figure}

In this section we will discuss the different sources of gamma radiation
resulting from dark matter decay. Generically, $\gamma$-rays are expected as a
secondary product of the interactions of CRs with the galactic gas and
radiation field.  It is well known that CRs in the energy range we are
interested in obey the following  propagation equation (Ginzburg \&
Syrovatskii~\cite{gin1964})
\begin{eqnarray}
\label{eq:diffusion_equation} 
\fl\hspace{1cm}\frac{\partial N^{e^{\pm}}}{\partial t}
- {\bm \nabla}\cdot \left( D \,{\bm \nabla}
-\bm{v}_{c}\right)N^{e^{\pm}} +
\frac{\partial}{\partial p}
\left(\dot{p}-\frac{p}{3}\bm{\nabla}\cdot\bm{v}_{c}\right)
N^{e^{\pm}}-\frac{\partial}{\partial p} p^2 D_{pp} \frac{\partial}{\partial
p} \frac{N^{e^{\pm}}}{p^2} =  \nonumber \\ \hspace{2cm}
=  Q^{e^{\pm}}(p,r,z) + c\,\beta\,
n_{\rm gas}(r,z)\, \sigma_{e^{\pm}}\,N^{CR}\;.
\end{eqnarray}
Here $N^{e^{\pm}}(p,r,z)$ is the number density of the electrons/positrons;
$p$ is their momentum; $\beta$ their velocity in units of the speed of light
$c$; $\sigma_{e^{\pm}}$ is the production cross section of electrons/positrons
from spallation of CR protons and heavier nuclei with density $N^{CR}$ onto
the ISM gas, whose density is $n_{\rm gas}$; $D=\beta D_0 (\mathcal{R}/{\rm
GV})^\delta$ is the spatial diffusion coefficient which is a function of the
particle rigidity $\mathcal{R}$; $\bm{v}_{c}$ is the convection velocity;
$\dot{p}$ are continuous energy losses, whose relevant timescales $\tau =
p/\dot{p}$ are shown in Fig.~\ref{fig:energylosses}. The last term on the
l.h.s.~of Eq.~(\ref{eq:diffusion_equation}) describes diffusive reacceleration
of CR in the turbulent galactic magnetic field. In the quasi-liner theory the
diffusion coefficient in momentum space $D_{pp}$ is related to the spatial
diffusion coefficient by the relationship (see e.g.~\cite{dpp94})
\J{
\begin{equation}
D_{pp} = \frac{4}{3 \delta (4 - \delta^2)(4 - \delta)}
\frac{v_A^2~p^2}{ D}
\end{equation}
where  $v_A$ is the Alfv\`en velocity. 
Finally, $Q^{e^{\pm}}(p,r,z)$ represents the primary injection term.}

\J{The electron distribution can be computed numerically in the stationary limit $\partial N^{e^\pm}/\partial t \equiv 0$ by using the methods of~\cite{Strong:1998pw,Moskalenko:1997gh,Evoli:2008dv}.
However, for this paper we have used our own propagation codes, which have strongly improved convergence behavior
for mono-energetic injection spectra.
From this distribution we will then compute the $\gamma$-ray spectra and maps produced by ICS and bremsstrahlung. }

We will firstly discuss inverse Compton scattering and bremsstrahlung
radiation, and secondly prompt radiation, both with their galactic and
extragalactic contributions, with particular emphasis on the related
uncertainties.

\subsection{Inverse Compton scattering and bremsstrahlung}
Electrons and positrons produced in the decay of dark matter give rise to a
$\gamma$-ray signal, coming from the inverse Compton scattering (ICS) of low
energy photons of the interstellar radiation field (ISRF). A further, mostly
subdominant, contribution to the $\gamma$-ray flux comes from bremsstrahlung
of the electrons and positrons when scattering with the galactic gas.  
\J{In the context of the PAMELA positron excess,} ICS radiation from decaying dark matter
has been discussed in Refs.~\cite{Ishiwata:2009dk, Ibarra:2009nw}, for the
case of annihilating dark matter see Refs.~\cite{Cirelli:2009vg, Profumo:2009uf}.

\J{The ICS $\gamma$-ray flux due to injected mono-energetic electrons and}
positrons from dark matter decay inside of our Galaxy, and arriving on
Earth from the direction $\hat{\Omega}$ (characterized by galactic latitude
$b$ and longitude $l$), is given by
\begin{equation}
J_{\rm ICS}(\hat{\Omega},E_\gamma;E_e) = \frac{1}{4 \pi}\int_{\rm l.o.s.} ds
\int_{m_e}^\infty dE \, P(E_\gamma, E) \, n_e({\bf r},E; E_e) \;, 
\end{equation}
were $n_e({\bf r},E;E_e)$ satisfies the stationary solutions to the
propagation Eq. (\ref{eq:diffusion_equation}) for a source term with
monochromatic injection of positrons \J{and electrons at an energy} $E_e$, $P(E_{\gamma},E)$
is the differential photon emissivity and the spatial integral is along the
line of sight.  
\J{The relevant source function is given by
\begin{equation}
\label{eq:DMsource}
Q(\bm{r},E)=\frac{\rho_\chi(\bm{r})}{\tau_\chi m_\chi}\,\delta(E-E_{e})\;.
\end{equation}
where  $m_\chi$ is the DM particle mass and $\tau_\chi$ is its lifetime.}

As dark matter profile $\rho_\chi({\bf r})$ we take (assuming, as customary,
spherical symmetry) the NFW profile~\cite{NFW}
\begin{equation}
\rho_\chi^{\rm NFW}(r) \propto \frac{r}{r_s}\frac{1}{(1+r/r_s)^2}\;,
\label{eqn:NFWprofile}
\end{equation}
which we normalize to $\rho_\odot=0.36\GeV/\cm^3$ at the position of the Sun,
$r_\odot = 8.5\kpc$, and we adopt $r_s=20\kpc$. Uncertainties from the halo
profile will be discussed below.

The differential emissivity $P(E_\gamma, E) $ corresponds to two processes,
\begin{equation}
P(E_\gamma, E) =P_{\rm IC}(E_\gamma, E) + P_{\rm bremss}(E_\gamma, E) \ .
\end{equation}
The first term $P_{\rm IC}(E_\gamma, E)$ corresponds to inverse Compton
\J{scattering, which is derived by convolving the differential number density of
target photons with the well known Klein-Nishina cross section.} 
The second term, $P_{\rm bremss}(E_\gamma, E)$, stems
from the bremsstrahlung emission due to deflection of relativistic electrons
and positrons in the electrostatic potential of interstellar gas atoms and
molecules. Since this contribution is subdominant in the energy range we
consider in this work, we will always refer to the ICS channel as our
reference channel in the following. More details of the calculation can be
found in \cite{icsbrm}.

The \textit{galactic radiation field} in the galactic plane consists mainly of
starlight at wavelength \J{$\lambda\sim 1~\mu{\rm m}$, diffuse dust emission at
$\lambda \sim 100~\mu{\rm m}$} and the CMB. Notice however that since most
of the radiation (CMB aside) is produced in the galactic plane, the actual
radiation fields at large latitudes are poorly known, and the computation of
ICS radiation from high latitudes suffers from large uncertainties. 
In this work we will adopt the ISRF presented in Ref.~\cite{Porter:2005qx}.

\J{The} predicted $\gamma$-ray flux crucially depends on the
\textit{propagation model}. We list in Tab.~\ref{tab:prop_model} three
different models characterized by different choices of the propagation
parameters in Eq.~(\ref{eq:diffusion_equation}). \green{By comparing the results obtained with these three models we will estimate the uncertainties  associated to the poor knowledge of the propagation models.}  The MIN and MAX models are diffusion-reacceleration-convection models. They were proposed in
\cite{Donato:2001ms,Donato:2003xg, Maurin:2002hw} and are known to be
compatible with the observed B/C ratio and produce the minimum and maximum \J{observationally allowed}
fluxes of antiprotons on Earth, respectively.  The L1 model is derived from
\cite{DiBernardo:2009ku}, in which an energy dependent analysis of recent data
about secondary/primary ratios allowed a fairly accurate study of the
diffusion parameters $\delta$ and $D_{0}/L$.\footnote{It is a general fact
that stable secondary/primary ratios do not allow to probe separately the
magnitude of the diffusion coefficient and the height of the diffusion region.
Unstable/stable ratios, such as the $^{10}$Be/$^{9}$Be, can in principle
provide such a discrimination. The available data on such ratios, however,
have very large errors, so that it is extremely difficult to extract
information from them.} Although the main aim of \cite{DiBernardo:2009ku} was
not to find a best fit model able to reproduce all the observed spectra, it is
remarkable that the diffusion parameters determined via an high energy
analysis are able to describe data down to energies of the order of 1 GeV/\J{nucleon},  
and also to reproduce \J{with reasonable accuracy the} PAMELA \J{measurements of the antiproton flux}. 
\green{It must be noticed, however, that this is only one out of many different possible models in agreement with nuclei CR observations (also simple leaky-box models can in fact succeed in reproducing high energy data \cite{gaisser1992}). In fact, all three models listed in Tab.~\ref{tab:prop_model} well reproduce available data, in spite of the different parameters they adopt. Moreover, all these models assume cylindrical symmetry and diffusion to be homogeneous and uniform over the whole Galaxy, which is a rough approximation of the galactic propagation regime. Yet, this simple modeling allows to reasonably reproduce available data with relatively few free parameters. A study of the effects of a possible spatial dependence of the diffusion coefficient in a general three-dimensional geometry is beyond the scope of this work.}


\begin{table}[tb]
\begin{center}
  \begin{tabular}{|c||c|c|c|c|c|c|c|c|}
    \hline
    Model  & $\delta^{1}$ & $D_0$      & $R$  & $L$ & $V_{c}$ &$dV_c/dz$   & $V_{a}$  & $ h_{\rm reac}$\\
    &                          & [$10^{28}{\rm cm}^2/{\rm s}$]&[kpc] & [kpc] & [km/s] &km/s/kpc     & [km/s] &[kpc]\\
    \hline
    MIN  & 0.85/0.85 &  0.048 &20 &1  & 13.5 &0 &  22.4 & 0.1\\
    L1   & 0.50/0.50 &  4.6     &20 &4  &0 &  0 & 10    &4           \\
    MAX  & 0.46/0.46 &  2.31 &20 &15 &  5   &0 & 117.6 & 0.1\\
    \hline
  \end{tabular}
  \caption{Typical combinations of diffusion parameters that are consistent
  with an analysis of CR nuclei secondary/primary ratios. The MIN and MAX
  propagation models correspond  to minimal and maximal primary antiproton
  fluxes, respectively, while the L1 model can provide a good description of
  B/C, $\bar{p}/p$ and data on other secondary/primary ratios above 1
  GeV/n.\\ $^{1}$ Below/above the break in rigidity at $\mathcal{R} =4$ GV for the
  MIN, MAX and L1 model.}
  \label{tab:prop_model}
\end{center}
\end{table}

\begin{figure}[tp]
\centering
\includegraphics[height=2.5 cm]{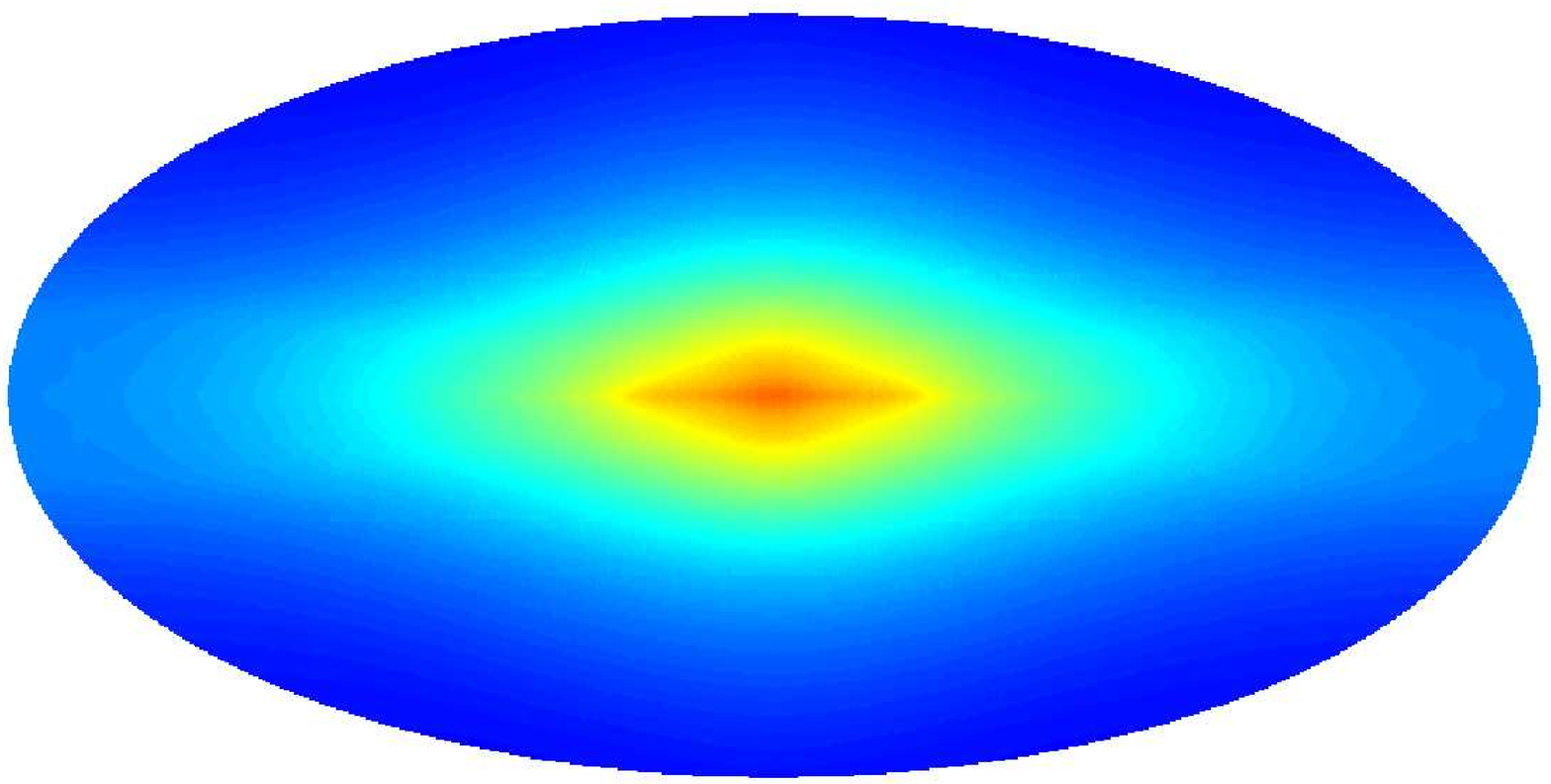}
\includegraphics[height=2.5 cm]{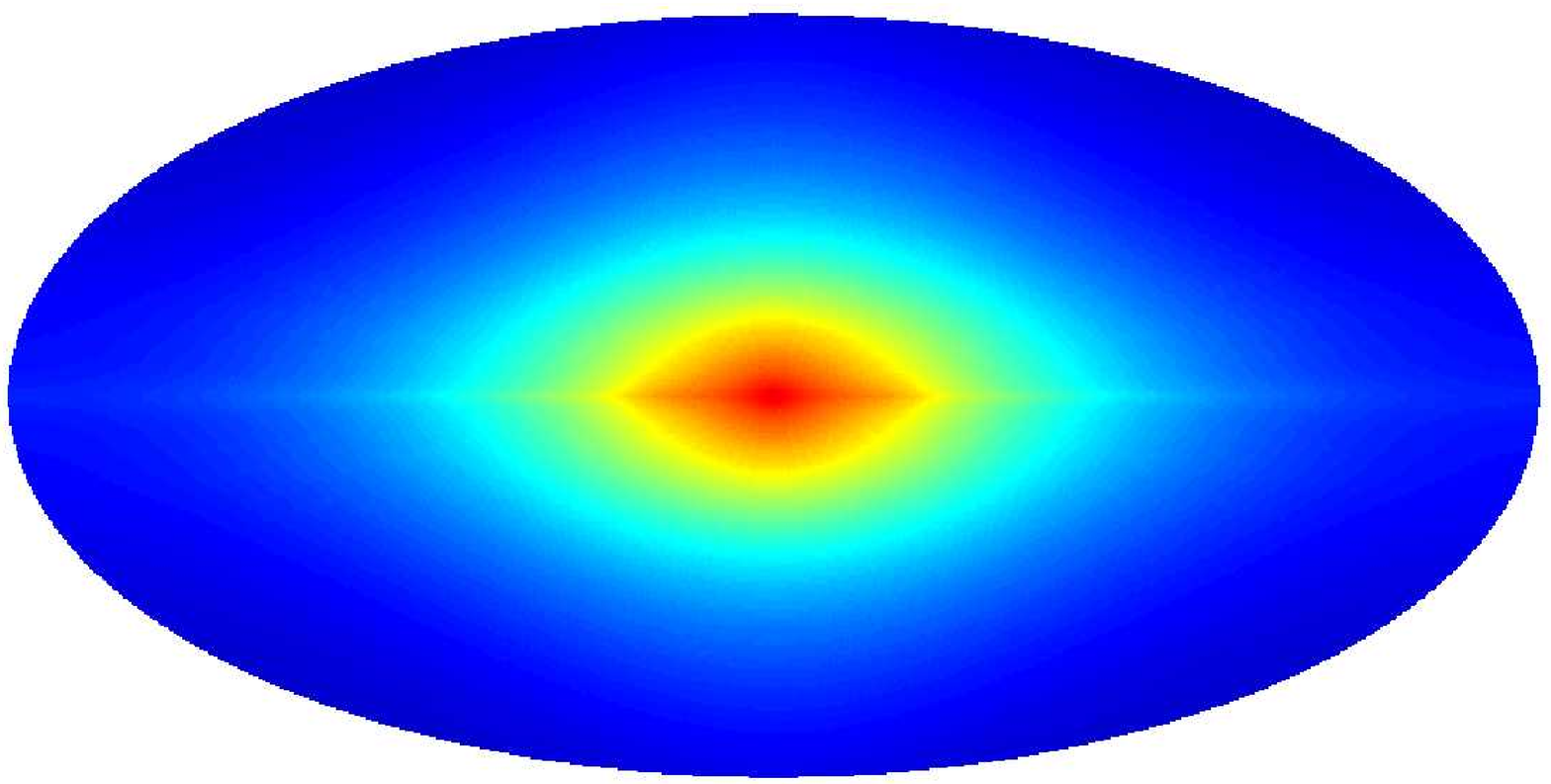}
\includegraphics[height=2.5 cm]{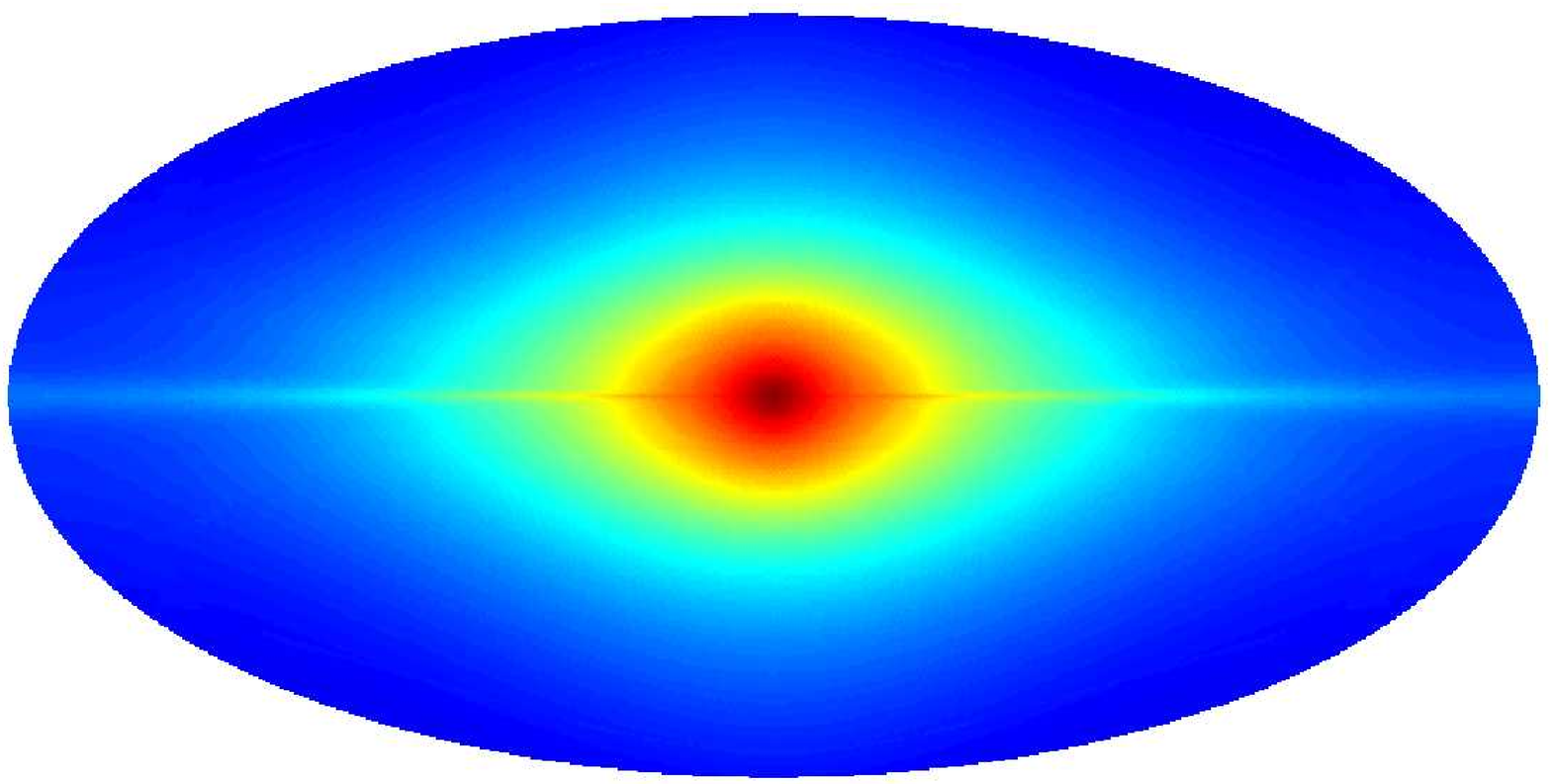}\\
\vspace{0.5cm}
\includegraphics[height=3.8 cm]{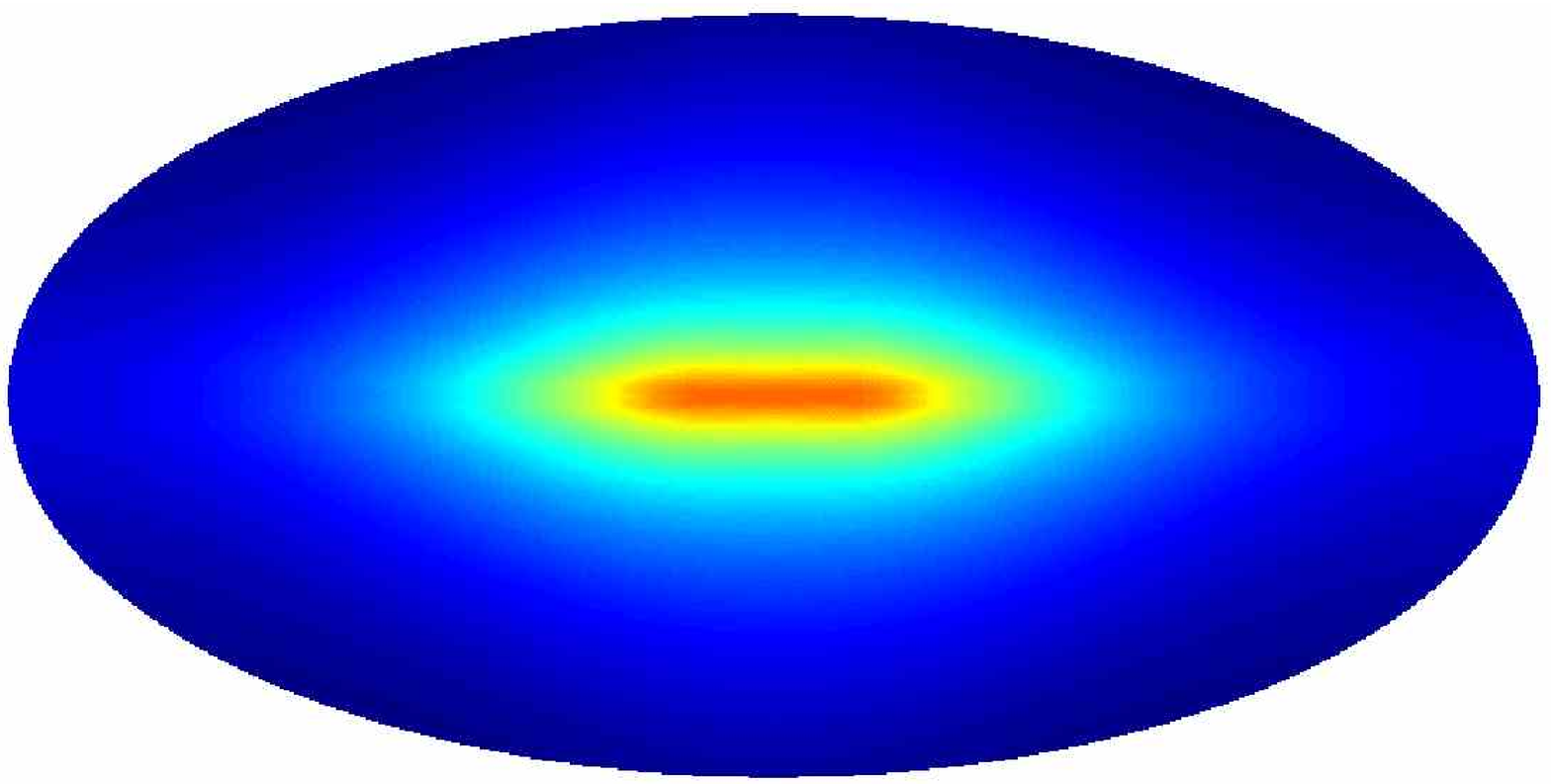}\\
\includegraphics[height=0.5 cm]{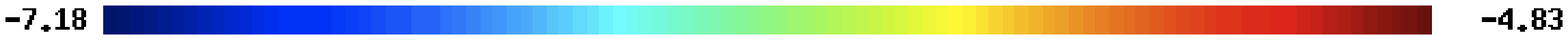}
\caption{The $\gamma$-ray emissions at 0.1 GeV, 1 GeV and 10 GeV (upper
panel from left to right) produced by dark matter particles decaying into
$e^+ e^-$ pairs, where $m_\chi=200\,$GeV, $\tau_\chi= 10^{26}\,$s.  Results
hold for the L1 diffusion model of Tab.~\ref{tab:prop_model} and for the NFW
halo profile. The lower panel shows the ICS radiation from astrophysical
sources at 10 GeV for comparison (again from model L1).  The color scaling
corresponds to the logarithm to the base 10 of the flux in
GeV/s/$\rm{cm}^2$/sr. Note that the color scale corresponds to the same flux
range in all panels.}
\label{fig:dm_prop_models}
\end{figure}

For illustrative purposes, we show in Fig.~\ref{fig:dm_prop_models} the
predicted $\gamma$-ray emission for the L1 model of Tab.~\ref{tab:prop_model}
at 0.1, 1 and 10 GeV, respectively, assuming a dark matter decay into $e^+
e^-$, where $m_\chi=200\,$GeV and $\tau_\chi=10^{26}\,$s. We also show for
comparison the ICS radiation from primary electrons of astrophysical origin.
In general, as apparent from these plots, dark matter induced ICS radiation
extends to higher latitudes than the ICS radiation due to the contribution of standard astrophysical source, which is mainly concentrated on the galactic disk. 

In addition to the ICS radiation produced in our Galaxy there is also a
related \textit{extragalactic} contribution, resulting from scattering of
electrons from dark matter decaying outside of our Galaxy with the CMB. This
component can potentially dominate the overall ICS fluxes at low energies, and
we include it for completeness. The calculation is straightforward and details
can be found in, \fex, Ref.~\cite{Ishiwata:2009dk,Profumo:2009uf}. In contrast
to these references, we also took into account absorption effects due to
inelastic scattering between ICS photons and the intergalactic background
light (IBL), following Ref.~\cite{Stecker:2005qs, Ibarra:2009nw} (adopting the
``fast evolution'' model).

Another contribution to the ICS radiation comes from electrons and positrons
produced in the dark matter decay inside of our dark matter halo, but outside
of the diffusion zone. If these particles are far enough from the galactic
disk region, beyond a few kpc, their main energy loss channel is scattering
with the CMB. The corresponding propagation length is $\mathcal{O}(100\kpc)$
for 1 TeV particles, and $\mathcal{O}(10\kpc)$ for 10 TeV particles.
Calculating the corresponding ICS flux while neglecting the motion of the
particles in general overestimates this flux by an $\mathcal{O}(1)$ factor.
Being conservative we do not include this radiation component to our bounds
and leave a more careful calculation for future work.

\paragraph{Uncertainties.} 
The largest uncertainties in the predicted $\gamma$-ray fluxes come from
poorly known propagation parameters, in particular from the height of the
diffusion zone.  The corresponding uncertainties can reach one order of
magnitude. 

As we already noticed, the height $L$ is only poorly known since its
determination is degenerate with the diffusion coefficient. The most widely
adopted range of variation of $L$ is between 1 and 15 kpc, based on the poor
quality data on $^{10}$Be/$^{9}$Be. Due to this uncertainty it is hard to
obtain any definitive constraint on, \fex, the supersymmetric parameter space
of DM based on current anti-proton data~\cite{Donato:2003xg, Adriani:2008zq}.
The forthcoming AMS-02 experiment \cite{DiFalco:2006mh} will however provide very accurate data on unstable/stable ratios (as well as for B/C and other stable
secondary/primary ratios), which might allow a more precise determination of
the diffusion height scale $L$.

In our case it is clear that a larger height of the diffusion zone leads to
more DM decays contributing to the $\gamma$-ray flux, because more electrons
of DM origin are confined in the diffusive region. Uncertainties on other CR
propagation parameters, such as the Alfv\`en velocity $v_{A}$ and the
convection velocity $v_{c}$, are less relevant in affecting the electron
distribution of DM, and especially affect only the electrons below around 10
GeV since higher energy electrons lose energy too rapidly via ICS and can not
propagate over long distances. 

We found that the full-sky ICS $\gamma-$ray emission induced by very high
energy electrons from DM decay obtained in the L1 model is comparable with the
one obtained using other widely known models, namely DC and
DR~\cite{Strong:1998pw} which adopt the same height of the diffusion zone as
our L1 model. 

Another source of uncertainty comes from the dependence of the $\gamma$-ray
emission on the halo profiles.  We compared the fluxes predicted for different
halo models (with parameters as in Ref.~\cite{Ibarra:2009nw}) and found that
for shallower halo profiles such as the Kra~\cite{Kra} and the
isothermal~\cite{Bergstrom:1997fj} profile the $\gamma$-ray emission is
reduced by around 10\%. On the other hand, in case of the Einasto profile (see
\cite{Pieri:2009je} and references therein) the flux is enhanced by 30\%.
Since this variation is subdominant when compared with the uncertainties of
the propagation models, we will simply adopt the NFW halo profile in the rest
of the present work.

The $\gamma$-ray emission also depends on the Galactic magnetic field, since
synchrotron losses can be of the order of ICS radiation losses at high
electron energies. The magnetic field profile close to the Galactic center is
quite uncertain and could be considerably higher than a few $\mu\rm
G$~\cite{melia}. \J{In this paper, }we adopt the field from Ref.~\cite{Strong:1998fr} which
matches the 408 MHz synchrotron distribution. We compared our results with
another widely used magnetic field  and found that the $\gamma$-ray fluxes at high energies increase by just 15\% by changing the magnetic field model to the one presented in Ref.~\cite{heiles96} which is based on a large-scale data set on starlight polarization.

\subsection{Prompt Radiation}
\J{Although this work focuses on ICS radiation produced by electrons and positrons
from decaying dark matter, we cannot overlook the fact that the prompt
radiation of gamma rays can be a very competitive signature of dark matter decays.
Indeed, in many realistic cases it turns out that this component can be larger than the ICS and therefore 
it can actually give the strongest constraints. }

Prompt radiation from dark matter decay is produced \J{more frequently inside our overdense} Galactic dark
matter halo\footnote{Although the halo profile is expected to be approximately isotropic, the corresponding
flux at Earth exhibits a strong dipole-like anisotropy due to the offset
between the Sun and the galactic center, which can be used to distinguish it
from the extragalactic $\gamma$-ray background, see Ref.~\cite{Ibarra:2009nw}.}
but it is also produced at cosmological distances. 
At energies around $10\GeV$ or below, the magnitude of the halo and extragalactic fluxes are of
the same order when looking in direction of the anti-galactic center, whereas
at higher energies around and above $100\GeV$ the inelastic scattering between
$\gamma$ rays and the IBL reduces the extragalactic component significantly
and can \textit{not} be neglected.

We include the galactic and the extragalactic prompt radiation, following the
calculations outlined in Ref.~\cite{Ibarra:2009nw}. Uncertainties come mainly
from the adopted dark matter halo profile and its normalization at position of
the Sun.

%
%

\section{$e^\pm$-response Functions from Inverse Compton Scattering} 
In this section we derive the $e^\pm$-response functions for the $\gamma$-ray
emission induced by DM decays into electrons and positrons by comparing our
predictions with the 1-year observations of \J{Fermi LAT in the energy range of 0.5 GeV to 300 GeV. }

As data we use the Fermi LAT $\gamma$-ray maps as derived in
Ref.~\cite{Dobler:2009xz}.\footnote{We also performed an analysis of the
publicly available event data on \url{http://fermi.gsfc.nasa.gov/ssc/data/},
coming to identical results as Ref.~\cite{Dobler:2009xz}. We decided to use
the maps from~\cite{Dobler:2009xz} in order to keep the data basis of our analysis
easily accessible.} In this analysis the events were binned into the eight
energy ranges 0.5 - 1 GeV, 1 - 2 GeV, 2 - 5 GeV, 5 - 10 GeV, 10 - 20 GeV, 20 - 50 GeV, 50 - 100 GeV and 100 - 300 GeV.

A few words of caution are required concerning the data basis: The adopted
$\gamma$-ray maps are based on the ``diffuse'' event class, which at energies
above 50 or 100 GeV suffers from background contamination \cite{AckermannTalk}
that becomes relevant when diffuse fluxes are studied. In the highest energy
regime the background contamination might be on the 50 - 80\% level.
Furthermore, no attempt of subtracting point sources were made. The adopted
statistical errors are always derived from the exposure $3\times 10^{10} {\rm
cm}^2\, {\rm s}$, which is good enough for the purpose of this
paper.\footnote{We cross-checked with our own analysis of the publicly
available event data that this gives indeed the correct number counts at high
energies.} We do not attempt to include systematic errors in the analysis,
which in light of the large background contamination of the data would be too
premature.

\label{sec:con-r}

\subsection{Definitions and Optimal Sky Patch}
Following the same \textit{Ansatz} we made for the case of synchrotron
radiation proposed in Ref.~\cite{Zhang:2009pr}, we introduce response
functions, which are functions of the electron injection energy and are
associated to $\gamma$-ray observations in a sky patch $\Delta\Omega$ and in
an energy band $E_0\leq E_\gamma\leq E_1$. The $e^\pm$-response functions are
defined as the ratio of the predicted $\gamma$-ray fluxes resulting from
decaying dark matter to the observed fluxes as
\begin{eqnarray}\label{eq:response}
F_\gamma^{E_0:E_1}(\Delta\Omega;E_e) &\equiv&
\frac{\int_{E_0}^{E_1}dE_\gamma\int_{\Delta\Omega}d\Omega
J_{\rm ICS}(\Omega,E_\gamma;E_e)}
{J_{\rm obs}^{E_0:E_1}(\Delta\Omega)+2\cdot\delta J_{\rm obs}}
\left(\frac{\tau_\chi}{10^{26}\, \rm s}\right) \left(\frac{m_\chi}
{100\,{\rm GeV}}\right)\;,
\label{eqn:DefResp}
\end{eqnarray}
where $J_{\rm ICS}(\Omega,E_\gamma;E_e)$ is calculated for the injection
spectrum Eq.~(\ref{eq:DMsource}) and we adopt the conservative attitude of
adding, in each energy and angular bin, to the central value of the observed
flux the associated $2\sigma$ error. These functions depend on neither
$\tau_\chi$ nor $m_\chi$ of Eq.~(\ref{eq:DMsource}), and constraints on a
given DM decay model can then be easily cast in the form           
\begin{eqnarray}\label{eq:constraint}
\int_{m_e}^{m_\chi} d E_e\,
F_\gamma^{E_0:E_1}(\Delta\Omega;E_e)\frac{dN_e}{dE_e}&\leq&
\left(\frac{\tau_\chi}{10^{26}\, \rm s}\right)
\left(\frac{m_\chi}{100\,{\rm GeV}}\right) \;,
\end{eqnarray}
where $dN_{e}/dE_{e}$ is the electron/positron spectrum obtained from DM decay
in some specific particle physics model. We stress again that the advantage of
the $e^\pm$-response function approach is that the $e^\pm$-response functions
are independent of the specific DM decay spectrum. The method is hence
directly applicable to any DM model, and, moreover, allows a discussion of the
typical characteristics of ICS radiation from DM decay in a model-independent
way.

The $e^\pm$-response functions depend crucially on the chosen patch
$\Delta\Omega$, which in the ideal case should cover the area with the largest
signal-to-background ratio, maximizing the $e^\pm$-response functions.
For the sake of clarity of our approach, and to avoid problems with
statistical bias related to adaptive methods (which are related to downward
statistical fluctuations and become severe if the statistics is low, see
Ref.~\cite{Papucci:2009gd}), we choose a fixed patch with large signal-to-background
ratio by inspection of the signal-to-background maps. 

\begin{figure}[t!]
\centering
\includegraphics[height=6.0cm]{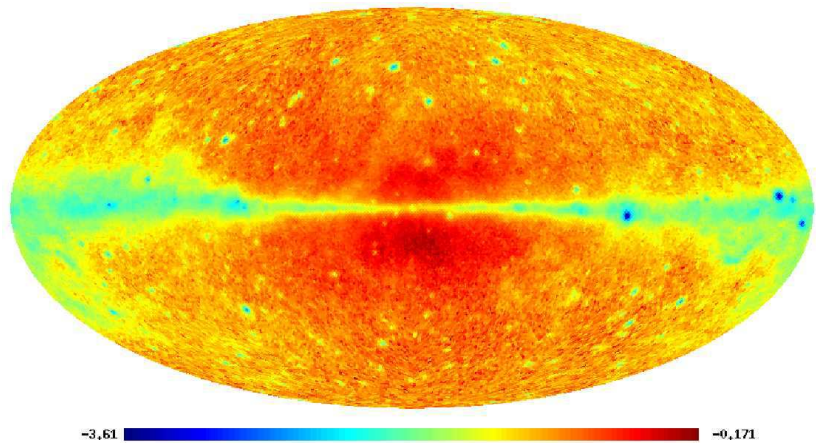}
\vspace{.5cm}
\caption{Signal-to-background map of ICS radiation from dark matter with
$m_\chi=200\,$GeV, $\tau_\chi=10^{26}\,$s decaying into $e^\pm$-pairs,
compared to the Fermi LAT $\gamma$-ray observations in the 0.5 - 1 GeV
regime. Results hold for the L1 propagation model of
Tab.~\ref{tab:prop_model}. Note the logarithmic color scaling, warmer colors
indicate larger signal-to-background.}
\vspace{.5cm}
\label{fig:excess}
\end{figure}

Clearly, the optimal region might in principle depend on the observed energy
range and injection energy of electrons and positrons. But it turns out that
for injection energies around 100 GeV - 10 TeV and high enough $\gamma$-ray
energies, the optimal region is always located south of the galactic center.
This situation is illustrated in Fig.~\ref{fig:excess}, which shows an example
of a signal-to-background map for observed energies between 0.5 and 1 GeV,
where the statistics is very good, for dark matter decay producing
monochromatic electrons and positrons at 100 GeV. The map already suggests
that the location of the optimal patch for constraining inverse Compton light
from decaying dark matter actually lies in a region close to the galactic
center, located south of the galactic plane. 

Before discussing this further we note that an exception occurs for very
high injection energies in the 1 - 10 TeV region, and low enough observed
$\gamma$-ray energies. There the overall ICS flux can actually be dominated by
extragalactic ICS contributions from scattering between electrons and
positrons from dark matter decay with the CMB. In these cases the optimal
patch would be located at the pole regions. The same holds in general true for
prompt radiation from dark matter decay in and outside of the Galaxy, which
has a much shallower angular profile than the galactic ICS component.

\begin{figure}[t]
\begin{center}
  \includegraphics[height=4.5 cm]{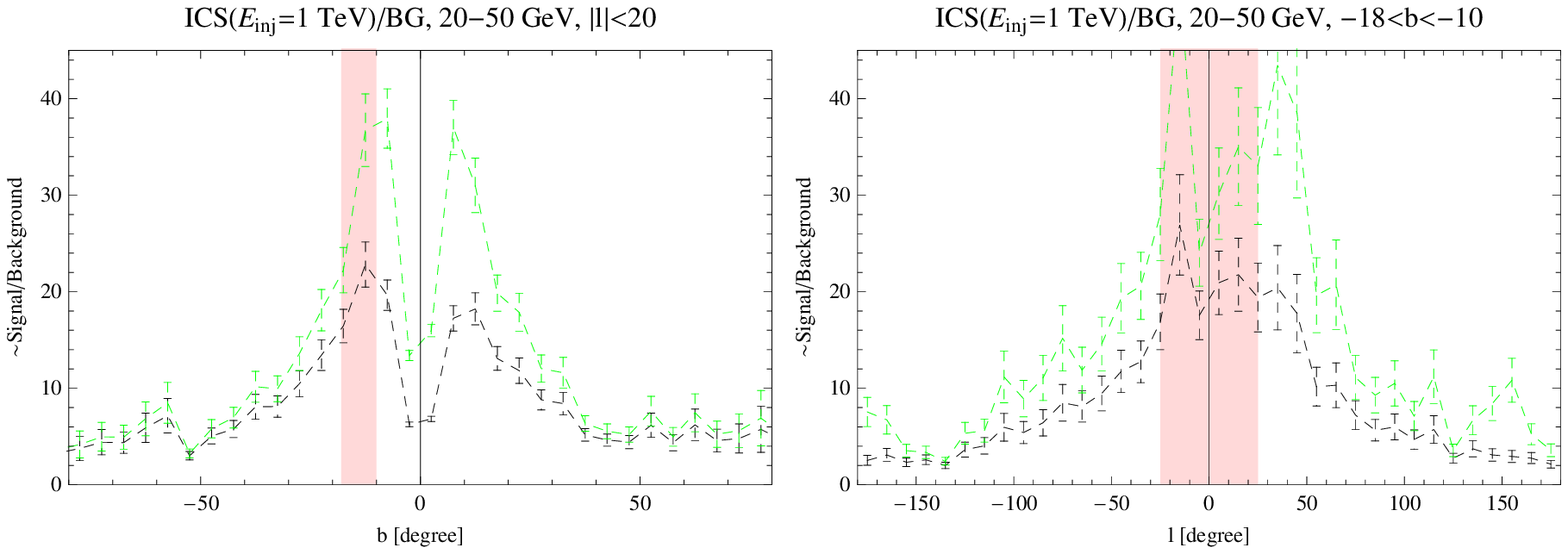}
  \includegraphics[height=4.5 cm]{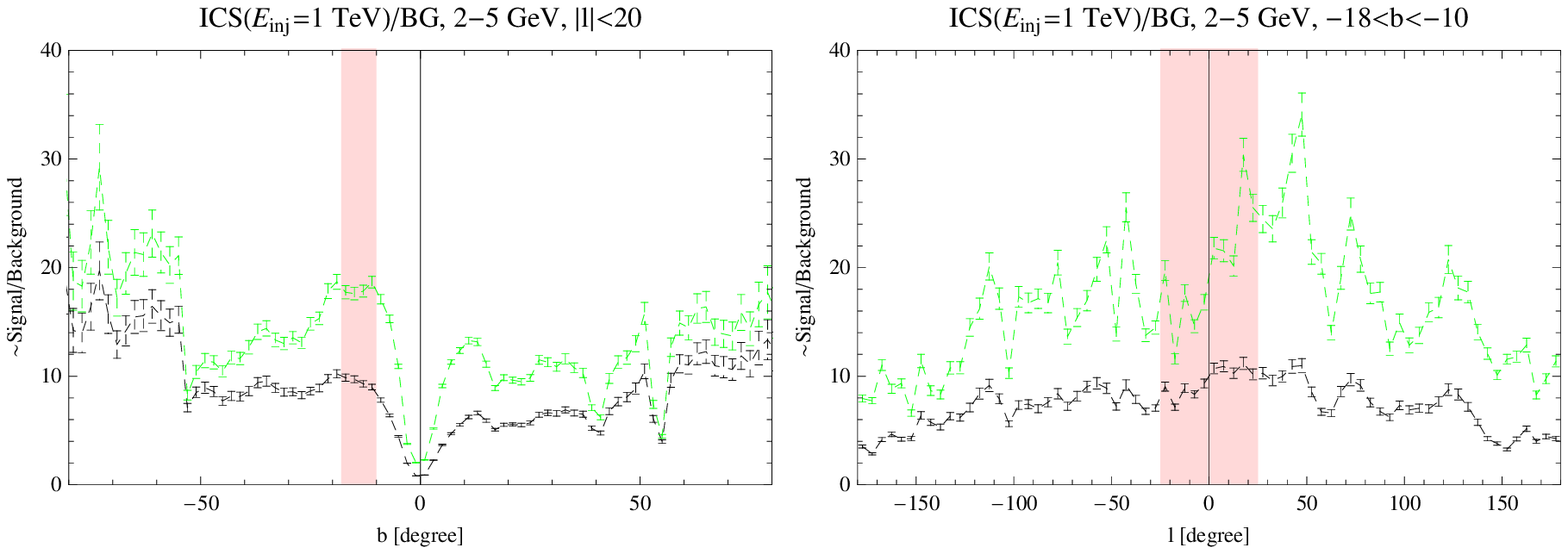}
  \includegraphics[height=4.5 cm]{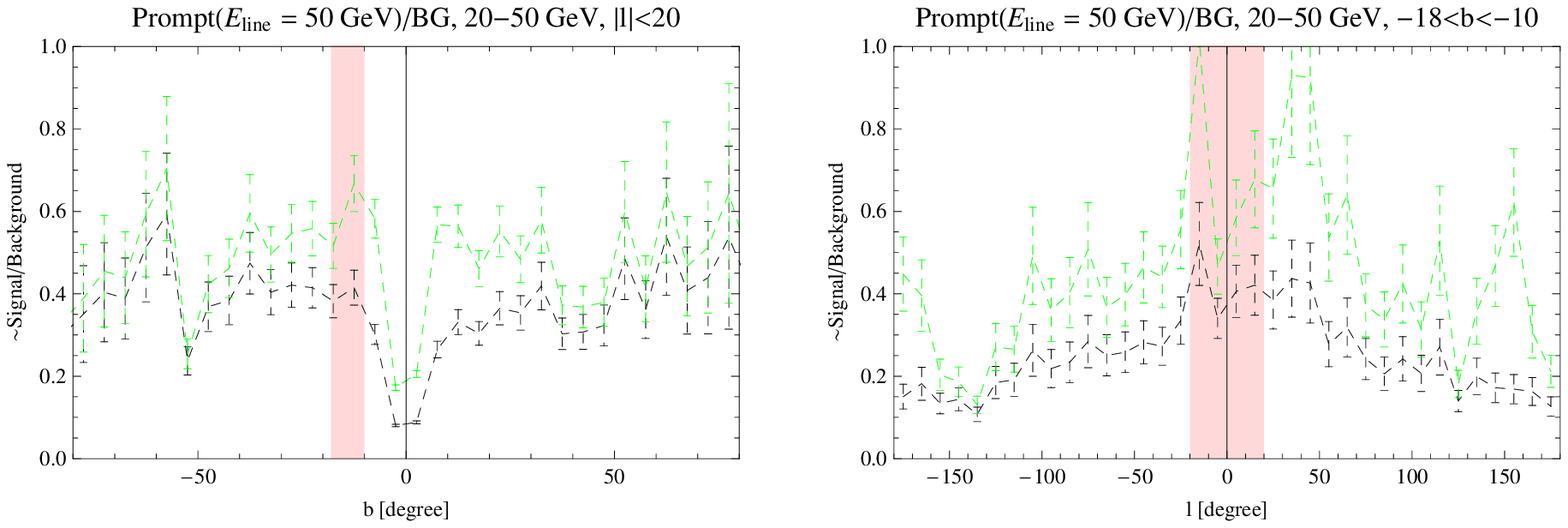}
\end{center}
\caption{Signal-to-background ratios as function of galactic latitude
(\textit{left panels}) and longitude (\textit{right panels}). The
\textit{upper and middle panels} correspond to pure ICS signal, the
\textit{lower panels} correspond to the pure prompt signal for comparison.
Extragalactic and galactic radiation are taken into account. The
\textit{black lines} take into account as the whole observed signal, the
\textit{green lines} are obtained after subtraction of our reference model
for the astrophysical component (Model L1). We find that the
signal-to-background ratio of ICS radiation at higher $\gamma$-ray energies
is maximized in the region~$\mathcal{S}$ defined by $|l|\leq20^\circ$ and
$-18^\circ\leq b\leq-10^\circ$, which is indicated by the light red shaded
region.}
\label{fig:SBplots}
\end{figure}

A more quantitative description of the situation can be found in the plots
shown in Fig.~\ref{fig:SBplots}. The black lines in the upper four panels show
the signal-to-background ratio in different observed energy ranges as function
of galactic latitude or longitude. The injection energy is now fixed to
$1\TeV$, but the results stay qualitatively the same for other injection
energies. As expected from the above discussion, at high energies
(\textit{upper panels}) the signal-to-background ratio is maximal in a region
close to the galactic center, whereas at very low energies (\textit{middle
panels}) it is maximal at high latitudes, due to the extragalactic ICS
component. The same dominance at high latitudes is also present in case of
prompt radiation (\textit{lower panels}).

For the derivation of the $e^\pm$-response functions we will concentrate on
the patch $\mathcal{S}$ close to the galactic center and defined by
$|l|<20^\circ$ and $-18^\circ<b<-10^\circ$, which is marked by the colored
region in Fig.~\ref{fig:SBplots}.  We checked that this patch indeed maximizes
the obtained constraints when varying the patch boundaries, except for the
very highest energy region (100 GeV - 300 GeV), where however the statistical
error is large. For this patch, we construct the $e^\pm$-response functions by
performing simulations injecting different mono-energetic electrons and
comparing the results with observations according to Eq.~(\ref{eqn:DefResp}).

Given that in some cases the
optimal patch is actually located at the galactic pole regions we will also
calculate and present bounds that come from comparing the preliminary
extragalactic $\gamma$-ray background as determined in Ref.~\cite{AckermannTalk}
with the extragalactic ICS and isotropic prompt radiation component, see below.

\begin{figure}[tp]
\centering
\includegraphics[height=8 cm]{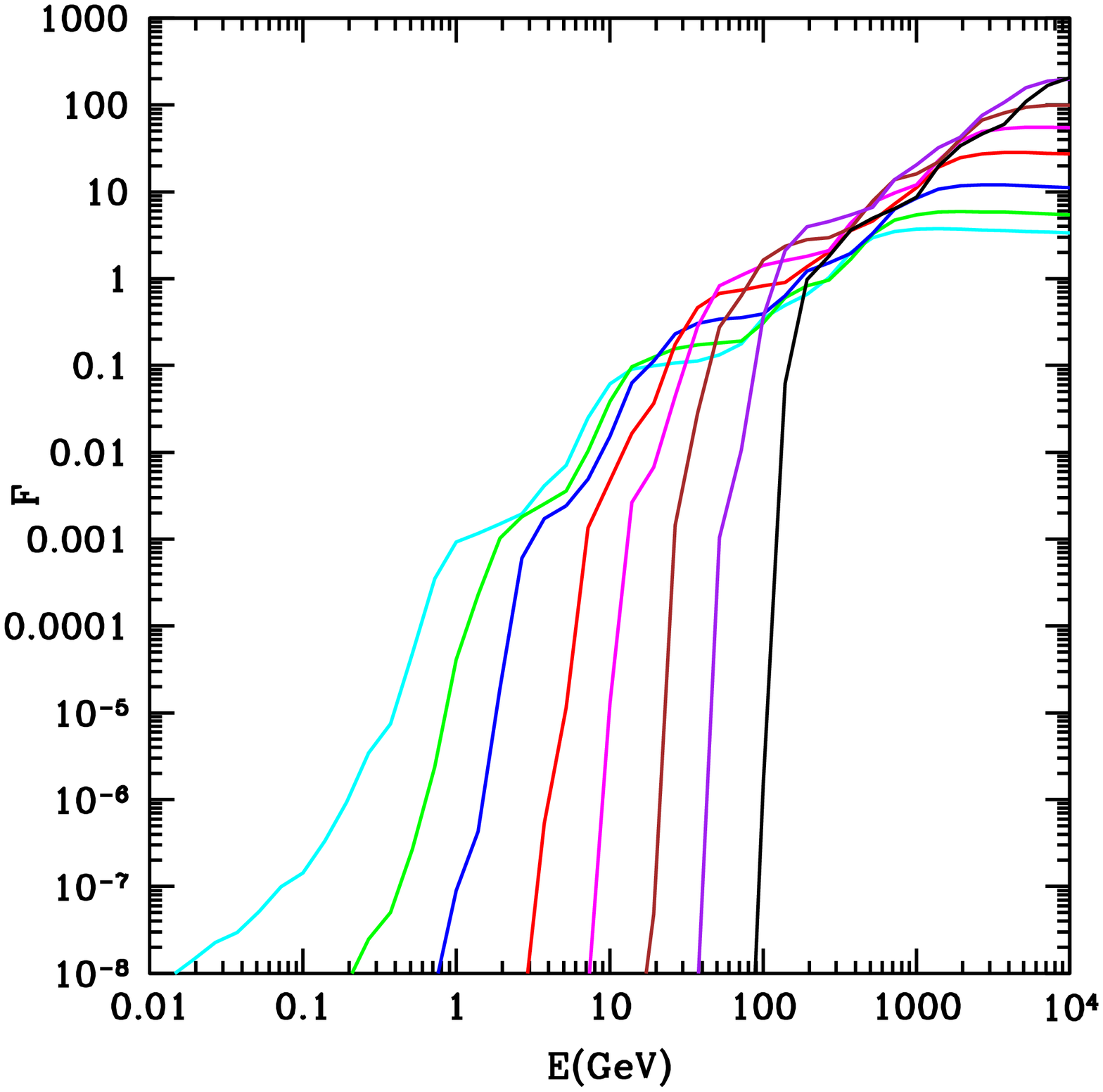}
\caption{The $e^\pm$-response function $F_\gamma$ based on $\gamma$-ray
emission for the L1 model of Tab.~\ref{tab:prop_model}. The $e^\pm$-response
functions are derived from the eight $\gamma-$ray energy ranges $0.5-1$ GeV,
$1-2$ GeV, $2-5$ GeV, $5-10$ GeV, $10-20$ GeV, $20-50$ GeV, $50-100$ GeV,
and $100-300$ GeV from top to bottom at left side, respectively. The
underlying sky patch~$\mathcal{S}$ is defined by $|l|\leq20^\circ$ and
$-18^\circ \leq b\leq -10^\circ $. 
} 
\label{fig:l1}
\end{figure}

\subsection{Response Functions without Foreground Subtraction}
Our results for the $e^\pm$-response function are shown in Fig.~\ref{fig:l1}
as function of the electron/positron injection energy, for the 8 different
energy ranges of Fermi LAT skymaps from Ref.~\cite{Dobler:2009xz}. The highest
energy range provides the strongest constraint on decaying dark matter with
very hard electron/positron energy spectrum.  However, for lower injection
energies in the 100 GeV$-$1 TeV region, several energy ranges give actually
roughly the same constraints.

\begin{figure}[tp]
\centering
\includegraphics[height=9 cm]{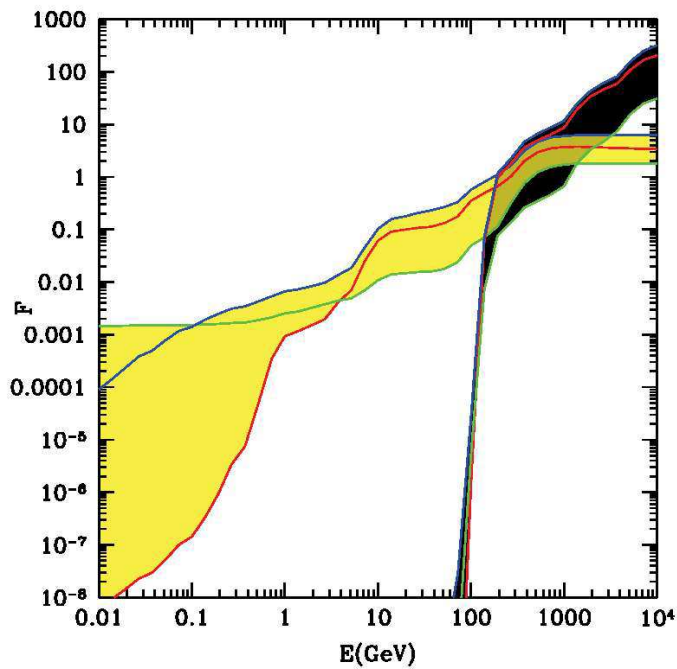}
\caption{The propagation model dependence of the $e^\pm$-response function
$F_\gamma$ based on our fixed patch for the $\gamma-$ray energy range
$0.5-1$ GeV (\textit{yellow band}, curves extending to low energies) and
$100-300$ GeV (\textit{black band}, curves cutting off around 100 GeV).  The
width of the bands represents the variation within the MIN (\textit{green}),
L1 (\textit{red}) and MAX (\textit{blue}) propagation models of
Tab.~\ref{tab:prop_model}.}
\label{fig:error}
\end{figure}

To illustrate the large uncertainties related to inverse Compton radiation
from dark matter decay inside the diffusive halo, we show in
Fig.~\ref{fig:error} the $e^\pm$-response functions based on the highest and
lowest $\gamma$-ray energy ranges for our three reference propagation models
from Tab.~\ref{tab:prop_model}. As emphasized before, the uncertainties on the
$e^\pm$-response functions are dominated by the propagation model, especially
for the lower energies, below 10 GeV injection energy, where also effects of
reacceleration become relevant. For higher injection energies above 10-100
GeV, where the response functions become of $\mathcal{O}(1)$ and are hence
relevant for the actual bounds, the uncertainties mainly stem from the height
of the diffusion zone.  In effect, high energy electrons and positrons lose
energy in a very short time compared to the diffusion time, thus making the
other details of the propagation irrelevant. The MAX propagation model gives
the strongest constraints due to its large diffusive halo, whereas the MIN
propagation model minimizes the constraints. Moreover, for the MAX model
re-acceleration shifts lower energy electrons to higher energies. This effect
is however only relevant for electrons below around 10 GeV, and thus increases
the $\gamma$-ray emission only in the MeV regime.  Note that for the highest
observed $\gamma$-ray energy region (100-300 GeV), one clearly finds a 
\J{sharp} cut off at low injection energies since $\gamma-$rays at such high
energies cannot be produced from ICS of electrons/positrons injected at
energies lower than 100 GeV. 


\subsection{Response Functions with Subtraction of Astrophysical Foregrounds}
\begin{figure}[tc]
\centering
\includegraphics[height=8 cm]{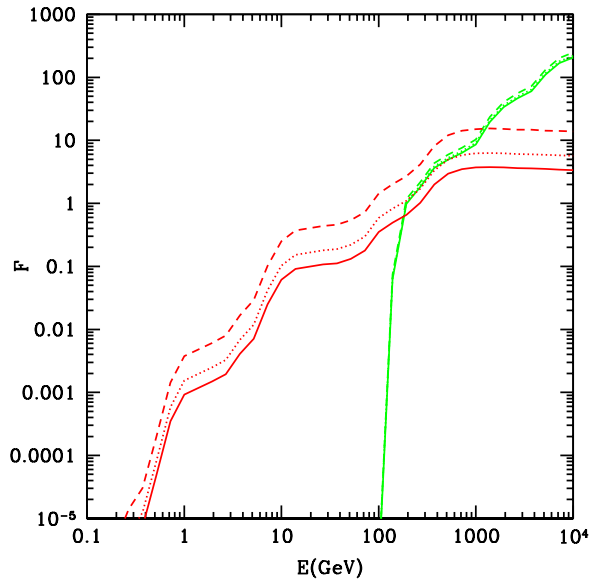}
\caption{The dependence of the $e^\pm$-response functions on subtraction of
astrophysical contributions to the $\gamma-$ray signal. The $e^\pm$-response
function for the L1 model based on the raw observed map (solid) and on
residual maps with $\gamma-$rays from $\pi^0$ decay (dotted) and from all
astrophysical processes (dashed, see text) removed. Red lines extending
below 1 GeV are based on $\gamma-$ray flux observed in the energy range
$0.5-1$ GeV and green lines are based on the interval $100-300$ GeV.}
\label{fig:sub}
\end{figure}

The $e^\pm$-response functions discussed so far are conservative because we
did not attempt to subtract any astrophysical contribution to the $\gamma$-ray
flux. In order to understand the conventional astrophysical $\gamma$-ray flux
one needs to estimate the $\gamma$-ray emission from different galactic
components. The most relevant production channels are nucleus-nucleus (mainly
proton-proton) photoproduction via $\pi^{0}$ decay and ICS and bremsstrahlung
of CR electrons and positrons.  It is generally found~\cite{Strong:1998fr} that
hadronically generated $\gamma$-rays dominate the flux at energies between 0.1
and 100 GeV and in the vicinity of the galactic plane, where most of the
interstellar gas is located, while at lower and higher energies and at high
latitudes ICS becomes comparable and can dominate. Bremsstrahlung is usually a
subdominant component.

In Fig.~\ref{fig:sub} we show the $e^\pm$-response function for the L1 model
based on residual $\gamma$-ray maps obtained by subtracting $\gamma$-rays
produced via $\pi^0$ decay, ICS and bremsstrahlung. In this foreground model
the electron flux is adjusted to always lie below the electron flux observed
by Fermi LAT, with a spectral index of around -3.2. The subtraction affects
the results at low energies. For example, at $E_e\sim10\,$TeV the
$e^\pm$-response functions based on $\gamma-$ray fluxes observed at energies
$0.5-1$ GeV are increased by a factor of around five when subtracting the total
astrophysical ``foreground'' at these $\gamma-$ray energies. Again at
$E_e\sim10\,$TeV the $e^\pm$-response functions based on $\gamma-$ray fluxes
observed at energies $100-300$ GeV are increased by $<10\%$ and $\sim15\%$
by the removal of $\gamma$-rays originating from $\pi^0$ decay and from all
astrophysical processes, respectively. This demonstrates explicitly that
constraints on dark matter decay can be improved by taking into account the
removal of astrophysical contributions mentioned above. At high energies the
removal turns out to be quite insufficient which is at least in part related
to the strong background contamination in the adopted data. This situation
will be improved when data with better background rejection becomes available.

%
%

\section{Constraint on Dark Matter Models}
\label{sec:constraint}

In the previous sections we constructed $e^\pm$-response functions for DM
decays which are independent of the particles physics details of the decay and
only depend on the spatial distribution of the DM particles and the
propagation of the produced electrons and positrons. These $e^\pm$-response
functions are very useful to estimate the constraints on DM decay based on
$\gamma$-rays produced by ICS within some specific DM model. However, within a
given DM model $\gamma$-rays can be produced not only as secondaries of
electron and positron propagation in the Galaxy, but also as final-state, or
prompt, radiation arising in the decay \cite{Bergstrom:2008gr,
Bergstrom:2004cy}. For selected decay channels, we calculated the
corresponding galactic and extragalactic prompt fluxes in our selected patch
as described above and added them to the one described by the
$e^\pm$-response functions in order to derive constraints
that come from the total prompt + ICS radiation flux of different dark matter
decaying models. Other works on constraints on the decaying/annihilating dark
matter interpretation of the PAMELA positrons excess with recent Fermi LAT
$\gamma$-data can be found in Ref.~\cite{Chen:2009uq, Papucci:2009gd,
Cirelli:2009dv}.

\begin{figure}[t]
\centering
\includegraphics[height=6.5 cm]{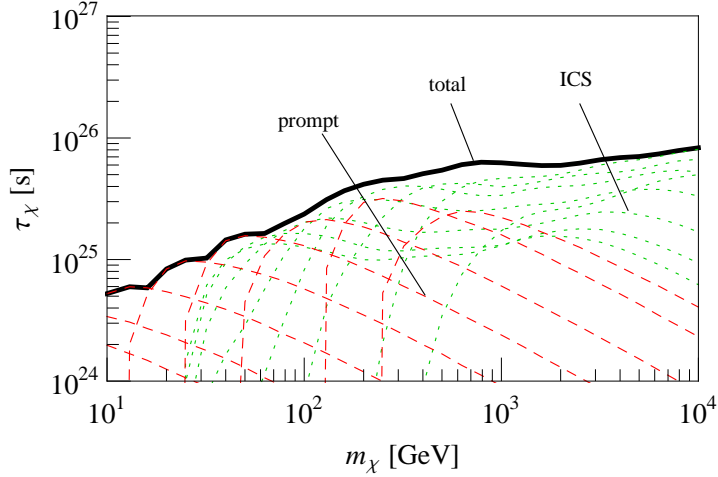}
\caption{Constraints on decaying dark matter for the decay channel $\chi\to
\mu^+\mu^-$ decoded into its different components. The thick solid line
shows the overall bounds on mass and lifetime, \textit{cf.}~also
Fig.~\ref{fig:bound}. Green lines represent the constraint coming from the
$e^\pm$-response function for ICS emission alone, whereas red lines are
based on the prompt photon spectrum alone. Each of the eight lines
corresponds to one of the observed $\gamma-$ray energy ranges as denoted in
the caption of Fig.~\ref{fig:l1}.}
\label{fig:boundD}
\end{figure}

To illustrate the interplay of prompt and ICS radiation bounds, we show them
in detail for the exemplary decay channel into $\mu^+\mu^-$ in
Fig.~\ref{fig:boundD}. In this plot, each of the green lines corresponds to
bounds coming purely from the galactic and extragalactic ICS radiation
(calculated from our response functions) for the eight different energy
regimes of the data. Regions below the green lines are excluded. On the other
hand, the red lines show the corresponding bounds when only prompt radiation
is taken into account. The thick black line is obtained when both radiation
components are combined for each energy regime separately. All fluxes are
calculated within our patch $\mathcal{S}$. As obvious from this plot ICS radiation bounds
dominate at dark matter masses above a few 100 GeV.\\

\begin{figure}[t]
\centering
\includegraphics[height=4.5 cm]{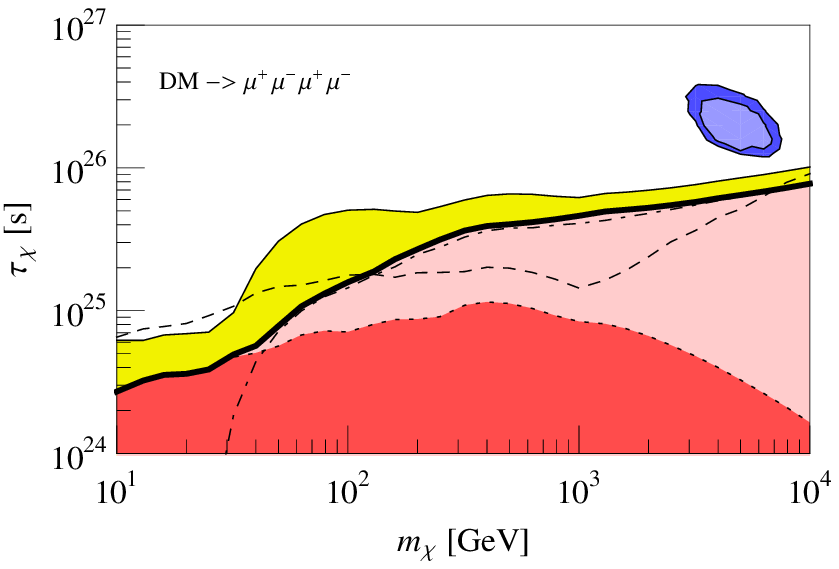}
\includegraphics[height=4.5 cm]{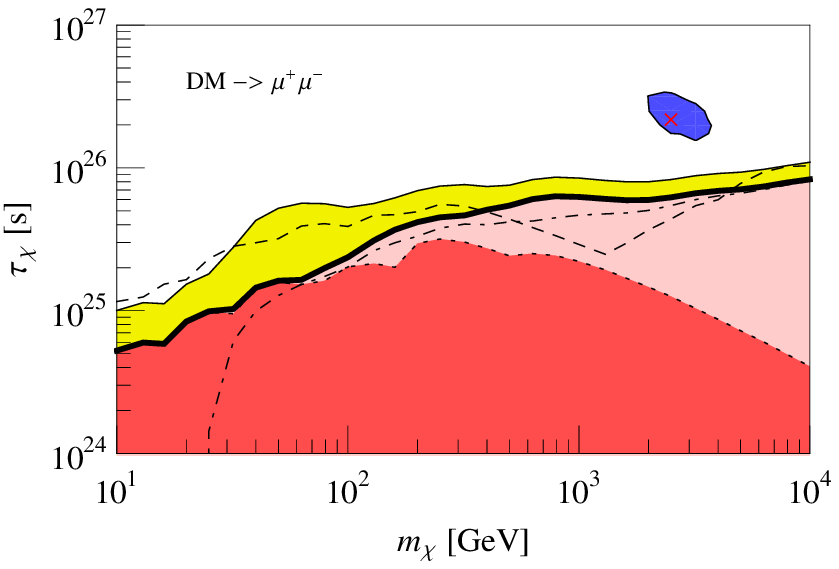}\\
\includegraphics[height=4.5 cm]{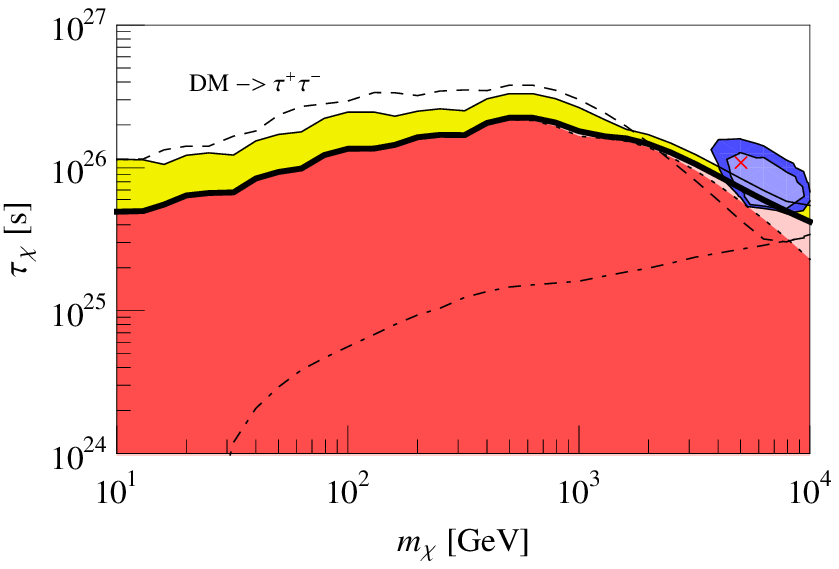}
\includegraphics[height=4.5 cm]{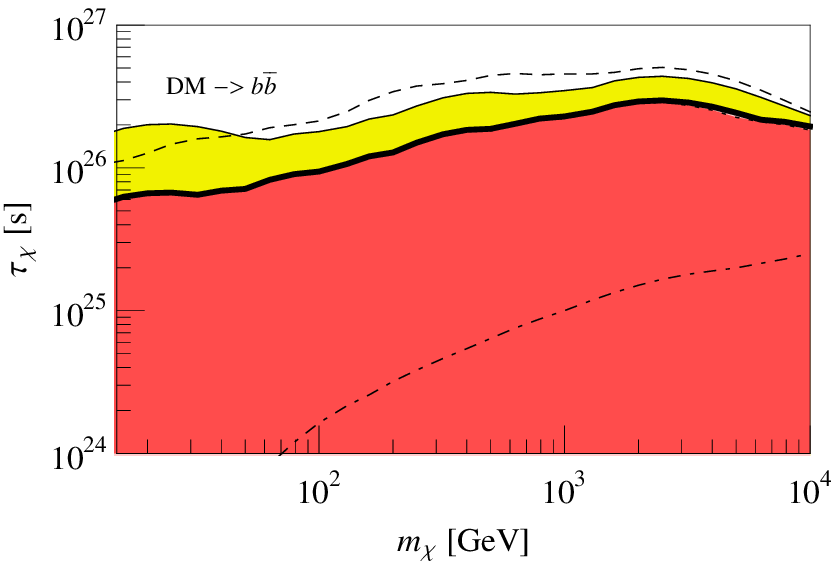}\\
\caption{Bounds on different decay channels in the mass
\textit{vs.}~lifetime plane. Regions below the \textit{thick solid} line
are excluded by combined ICS and prompt radiation in the L1 propagation
model, whereas parameter space below the \textit{dashed-dotted}
(\textit{dashed}) line is excluded due to ICS (prompt) radiation alone.
The ICS constraints shown with the \textit{dashed-dotted} lines are
calculated from the response functions shown in Fig.~\ref{fig:l1}. The
constraints can be strengthened to the yellow light shaded region if the
predictions of Model L1 for the galactic diffuse astrophysical foreground is
subtracted. The blue blobs and red crosses (which are taken from Ref.~\cite{Ibarra:2009dr}) show the parameters that well fit
electron + positron fluxes observed by Fermi LAT and HESS and the positron
fraction observed by PAMELA as described in the text.}
\label{fig:bound}
\end{figure}

In Fig.~\ref{fig:bound} we show our results for bounds on the four different
decay channels into $\mu^{+}\mu^{-}\mu^{+}\mu^{-}$, $\mu^+\mu^-$,
$\tau^+\tau^-$ and $b\bar{b}$ as examples with different amount of prompt and
ICS radiation.\footnote{The four-body decay into muons proceeds via two
intermediate neutral scalar particles with masses of 1~GeV.} The first three
decay modes can well fit the PAMELA/Fermi positron and electron data if the
positron excess is interpreted in terms of decaying dark matter, and the
preferred mass and lifetime regions are indicated by the blue blobs and red
crosses.\footnote{The shown regions should be understood as typical masses and
lifetimes that well fit the data. We performed a $\chi^2$ test, keeping the
electron background as a freely adjustable powerlaw with spectral index
between -3.3 and -3.0, whereas the positron background is kept fixed as the
one from ``Model 0'' in Ref.~\cite{Grasso:2009ma}. We include in the fitting
procedure only the PAMELA data on the positron fraction above
$10\;\textrm{GeV}$, as they should be less affected by solar modulation, and
the Fermi and HESS $e^\pm$-data, for which we added the corresponding
systematic and statistical errors in square.  Since systematic errors are
correlated, the obtained $\chi^2$ are relatively small, but in view of the
large uncertainties in the electron background a more detailed fit is not
reasonable.  The blue regions correspond to $\chi^2/\textrm{dof}=1$ and
$\chi^2/\textrm{dof}=0.75$, outside these regions the fits to HESS and Fermi
LAT become problematic.} We used the \textsc{Pythia}
package~\cite{Sjostrand:2000wi} to derive the electron, positron and
$\gamma$-ray decay spectra.

In these plots the dashed-dotted (dotted) line shows the bounds obtained from
ICS (prompt) radiation in our patch $\mathcal{S}$ alone, the thick solid line
shows the bounds obtained when prompt and ICS radiation are combined.
Furthermore, the bounds can be strengthened to the yellow region when the
foreground model L1 is subtracted from the data.

It turns out that for decays into $\mu^+\mu^-$ pairs and four-body decay into
$\mu^+\mu^-\mu^+\mu-$, the strongest constraints typically come from ICS
rather than from the prompt radiation and the constraints could be improved by
more than a factor of 2 for small masses and by a few 10\% for large masses
after removal of the $\gamma$-ray emission from conventional astrophysical
sources. In the case of decay into $\tau^+\tau^-$ and $b\bar{b}$ the prompt
radiation alone already provides strong constraints, which can again be improved by subtracting galactic foreground as for the case
of decay into muons. 


Note that our patch $\mathcal{S}$ is optimized for ICS radiation. Prompt
radiation from dark matter decay in general dominates at the galactic pole
regions, as discussed above (in the actual data, this behavior is disturbed at
high gamma-ray energies because of the large contamination of the data with
isotropic cosmic-ray background). Following the slicing of the sky as proposed
in Ref.~\cite{Papucci:2009gd}, we can find for the highest energy bin a patch
that actually increases our corresponding final state radiation bounds by
around 70\%.\footnote{Using this adaptively determined patch, which is located
at $10^\circ\leq b\leq20^\circ$ and $0\leq l \leq 10^\circ$ and has only a few
number counts, still does not allow to raise the bounds as high as shown in
Ref.~\cite{Papucci:2009gd}. The difference might originate in the smaller
energy bins used in~\cite{Papucci:2009gd}, and the inclusion of data above 300
GeV.}

For comparison, we also show with the dashed lines in Fig.~\ref{fig:bound} the
bounds that can be obtained by comparing the sum of extragalactic ICS
radiation, extragalactic prompt radiation and the maximal isotropic part of
the halo prompt radiation (which is identical to the flux from the Galactic anti-center) with the
preliminary results for the isotropic extragalactic gamma-ray flux as
presented in Ref.~\cite{AckermannTalk}. Comparing these bounds, which already
rely on a foreground subtraction, in case of decay into muons with the ones obtained from patch
$\mathcal{S}$ after foreground subtraction shows that they are subdominant and
become only relevant at very high masses. Our bounds are somewhat weaker
than the ones found in Ref.~\cite{Cirelli:2009dv}, which is due to our
inclusion of absorption effects and our more conservative treatment of
extragalactic ICS radiation.

\section{Conclusions}
\label{sec:conclusion}
In this study we have calculated the contribution to the $\gamma$-ray fluxes
from decaying dark matter particles, including the inverse Compton photons
resulting from energetic electrons and positrons through scattering with low
energy target photons in addition to the bremsstrahlung emissions. We
constructed $e^\pm$-response functions based on the full-sky $\gamma-$ray
observations by Fermi LAT which can be applied to constrain any decaying dark
matter model by convolving it with the specific decay spectrum into electrons
and positrons. We also studied the dependence of the $e^\pm$-response
functions on both the set of propagation parameters and halo profiles and find
that the most important uncertainty comes from the height of the diffusion
zone.

We applied the response functions to different decaying dark matter models
with leptonic final states, including also the effects of prompt radiation,
which can significantly increase the $\gamma-$ray emissions in the dark halo. 
Moreover, we demonstrated how the constraints can be further improved by
subtracting astrophysical contributions to the observed $\gamma-$ray flux.


Under the conservative assumption of a propagation model with the height of
the diffusion zone around 4 kpc, based on the Fermi LAT data, we can severely
constrain but not exclude models with dark matter decay into $\tau^+\tau^-$
that can explain the positron excess observed by PAMELA. Moreover, we find
that analogous models with two- and four-body decay channels into $\mu^\pm$s
remain essentially unconstrained by current observations. When our reference
foreground model is subtracted the lower bounds on the lifetime in general increase by $\mathcal{O}(1)$ factors for dark matter masses below 1 TeV, and by 10-60\%
for masses above 1 TeV, which is however not enough to exclude the above
channels in the parameter regime relevant for PAMELA. The bounds might improve
by $\mathcal{O}(1)$ factors when data with better background rejection is
used. For comparison we also calculated conservative bounds from the isotropic
extragalactic $\gamma$-ray background as inferred from the Fermi LAT data,
finding again that even the decay into $\tau^+\tau^-$ cannot be excluded in this
way.\\

In this work we demonstrated the use of $e^\pm$-response functions in
constraining dark matter models. We plan to update the response functions as
data with improved rejection of background becomes available.

\section*{Acknowledgements}
This work was supported by the Deutsche Forschungsgemeinschaft through SFB 676
``Particles, Strings and the Early Universe: The Structure of Matter and
Space-Time'' and through GRK 602 ``Future Developments in Particle
Physics''. LM acknowledges support from the State of Hamburg, through the Collaborative Research program ``Connecting Particles with the Cosmos'' within the framework of the LandesExzellenzInitiative (LEXI). CW thanks Alejandro Ibarra and David Tran for helpful discussions.


\end{document}